\documentclass
[aps,prl,twocolumn,superscriptaddress,showpacs,floatfix]{revtex4}%
\usepackage{graphicx}
\usepackage{dcolumn}
\usepackage{latexsym}
\usepackage{amssymb}
\usepackage{amsfonts}
\usepackage{amsmath}
\usepackage{fancybox}
\usepackage[binary,squaren]{SIunits}
\usepackage{colordvi}
\usepackage{color}%
\providecommand{\U}[1]{\protect\rule{.1in}{.1in}}
\providecommand{\U}[1]{\protect\rule{.1in}{.1in}}

\def\be{\begin{equation}}
\def\ee{\end{equation}}
\def\ber{\begin{eqnarray}}
\def\eer{\end{eqnarray}}

\begin{document}
\title{Laser-induced spatiotemporal dynamics of magnetic films}
\author{Ka Shen}
\affiliation{Kavli Institute of NanoScience, Delft University of Technology, Lorentzweg 1,
2628 CJ Delft, The Netherlands}
\author{Gerrit E. W. Bauer}
\affiliation{Institute for Materials Research and WPI-AIMR, Tohoku University, Sendai
980-8577, Japan}
\affiliation{Kavli Institute of NanoScience, Delft University of Technology, Lorentzweg 1,
2628 CJ Delft, The Netherlands}
\date{\today }

\begin{abstract}
We present a theory for the coherent magnetization dynamics induced by a
focused ultrafast laser beam in magnetic films, taking into account nonthermal
(inverse Faraday effect) and thermal (heating) actuation. The dynamic
conversion between spin waves and phonons is induced by the magnetoelastic
coupling that allows efficient propagation of angular momentum. The anisotropy
of the magnetoelastic coupling renders characteristic angle dependences of the
magnetization propagation that are strikingly different for thermal and
nonthermal actuation.

\end{abstract}

\maketitle



\textit{Introduction} --- Since the discovery of laser-induced ultrafast spin
dynamics in Nickel by Beaurepaire \textit{et al.}~\cite{Beaurepaire96}, the
spin manipulation in ferromagnetic system by femtosecond laser pulses has
attracted much attention since combining the intellectual challenge of new
physics with the application potential of ultrafast magnetization
reversal~\cite{Kirilyuk10}. Intense light can cause many effects in magnets,
such as the coherent inverse Faraday effect (IFE) as well as the excitation of
the coupled electron, magnon and phonon subsystems on various time scales. The
associated modulation of the magnetic anisotropy and magnetization modulus
allows coherent control of the magnetic order~\cite{Ju99,Kampen02}. The
transient magnetic field generated by the IFE allows non-thermal ultrafast
magnetization control~\cite{Kimel05,Hansteen05} that may be distinguished from
heating-induced effects by switching the light polarization. Nevertheless,
heating is essential for light-induced magnetization
reversal~\cite{Vahaplar09}. Toggle switching of the magnetization by heat
alone has also been reported~\cite{Barker13}. Understanding and controlling
the relative magnitude of thermal and non-thermal excitation is therefore an
important but unsolved issue.

\begin{figure}[ptb]
\includegraphics[width=6.5cm]{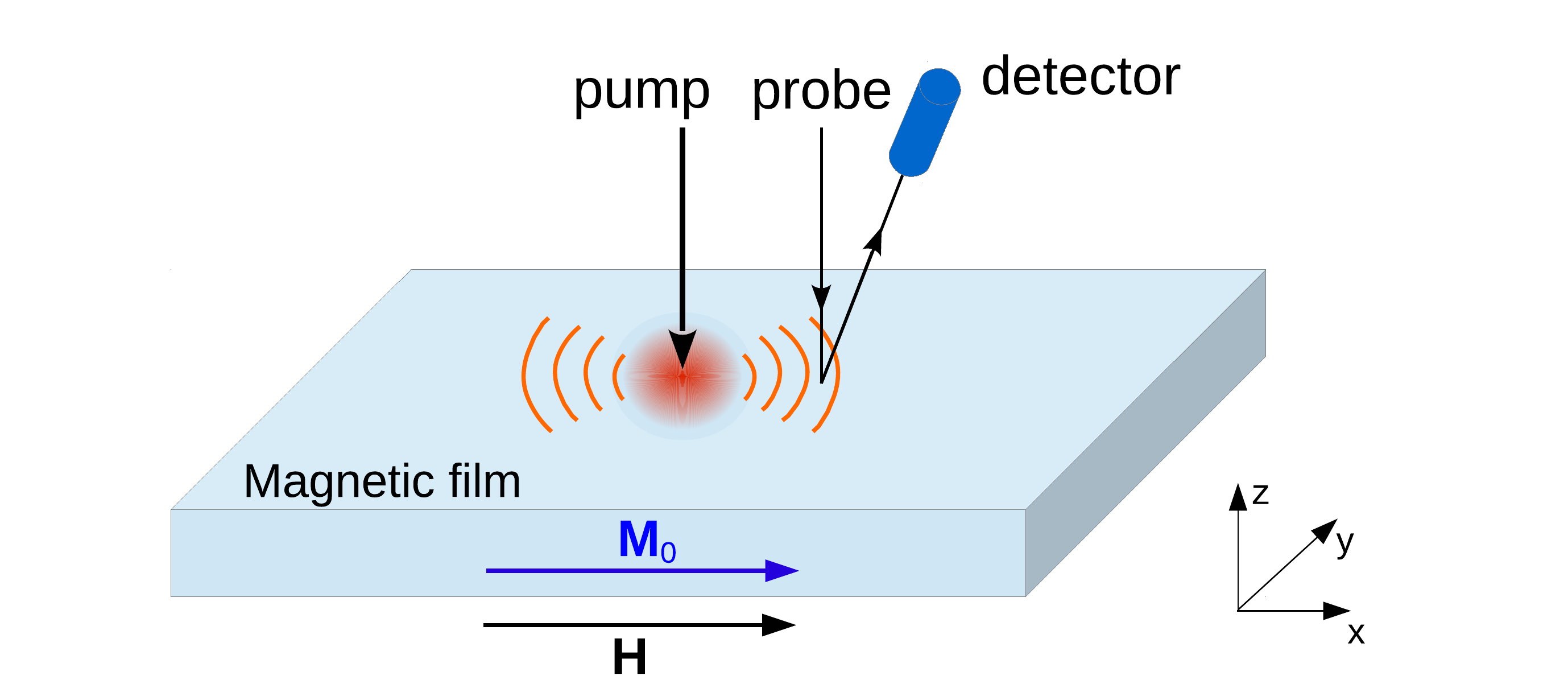}\caption{Pump-probe study of the
dynamics of magnetic films by pulsed lasers: An external magnetic field aligns
the equilibrium magnetization along the $x$ direction. The pump laser hits the
sample at the origin in time and space. The temporal distribution of the
out-of-plane ($z$) component of the magnetization is detected by the Kerr
rotation of the linearly polarized probe pulse.}%
\label{scheme}%
\end{figure}

The optical ultrafast pump-probe technique as shown in Fig.~\ref{scheme} is an
established powerful method to study matter.\ Here we will show that the
symmetry of the spatiotemporal magnetization distribution excited by a focused
laser beam reveals the relative contributions of thermal and non-thermal
excitations. This phenomenon originates from the magnetoelastic coupling
(MEC)~\cite{Kittel58,Akhiezer58,Schlomann60}, i.e., the coupling between spin
waves (magnons) and acoustic lattice waves (phonons). 

In the regions of phase space in which the magnon and phonon dispersion come
close, the MEC hybridizes magnons and phonons into coherently mixed
quasiparticles [\textquotedblleft magnon-polarons\textquotedblright(MPs)].
This implies that phonons in magnets can be converted into magnetization and
become detectable via magneto-optical~\cite{Scherbakov10,Kim12,Jager15} or
electrical~\cite{Uchida10b,Weiler12} techniques. In this Letter, we present a
study of the spatial magnetization dynamics in magnetic thin films after
focused-laser excitation \cite{Satoh12}. We consider here magnetic insulators
that are not affected by conduction or photo-excited free carriers. We find
that laser-generated phonons efficiently excite magnetization when the
diameter of the laser spot is comparable with the wavelength of the MPs. The
spatial dynamics of such phonon-induced magnetization shows a different
angular symmetry from that of the magnetization generated directly by laser
via the IFE. Dedicated experiments suggested here can therefore help
understanding the fundamental nature of light-matter interaction in magnets.

\textit{Formalism} --- The basic theory for MPs in special symmetry directions
by Kittel~\cite{Kittel58} and Akhiezer \textit{et al.}\cite{Akhiezer58}, extended by
Schl\"{o}mann~\cite{Schlomann60} to arbitrary propagating directions, was
developed more than half a century ago. The energy density of the minimal
model reads~\cite{Kittel58}
\begin{equation}
\mathcal{H}=\mathcal{H}_{\mathrm{ex}}+\mathcal{H}_{Z}+\mathcal{H}%
_{\mathrm{el}}+\mathcal{H}_{\mathrm{mec}}+\mathcal{H}_{\mathrm{dip}}.
\label{Hamil}%
\end{equation}
We adopt a cubic unit cell and consider the thin film limit in which the
magnetization is spatially constant over the film thickness. This assumption
holds for films up to ${100\,}\mathrm{\mu m}$~\cite{Satoh12}, which for
wide-gap insulators is still less than the penetration depth of the light and
allows us to use a two-dimensional model. With external magnetic field
$\mathbf{H}$ and equilibrium magnetization vector $\mathbf{M}_{0}%
\Vert\mathbf{x}$ ($\left\vert \mathbf{M}_{0}\right\vert =M_{0}$ saturation
magnetization) $\mathcal{H}_{\mathrm{ex}}=A[(\nabla m_{y})^{2}+(\nabla
m_{z})^{2}]$ and $\mathcal{H}_{Z}=-\mu_{0}HM_{0}+(\mu_{0}HM_{0}/2)(m_{y}%
^{2}+m_{z}^{2})$ represent the (linearized) exchange and Zeeman energies,
respectively, where $m_{y}$ and $m_{z}$ are the transverse magnetization
components of $\mathbf{m}=\mathbf{M}/M_{0}$. $\mathcal{H}_{\mathrm{el}}$ is
the lattice energy with both kinetic and elastic contributions $\mathcal{H}%
_{\mathrm{el}}=(1/2)\rho\dot{\mathbf{R}}\cdot\dot{\mathbf{R}}+(1/2)\lambda
(\sum_{i}S_{ii})^{2}+\mu\sum_{ij}S_{ij}^{2}$ with strain tensor $S_{ij}%
=(\partial_{i}R_{j}+\partial_{j}R_{i})/2$ and $\mathbf{R}$ representing the
lattice displacement with respect to equilibrium, $\rho$ the mass density, and
$\lambda$ and $\mu$ elastic constants. The MEC in Eq.~(\ref{Hamil}) reads
$\mathcal{H}_{\mathrm{mec}}=\sum_{i,j\in\{y,z\}}(b+a\delta_{ij})S_{ij}%
m_{i}m_{j}+2b\sum_{i\in\{y,z\}}S_{ix}m_{i}$,
where $a$ and $b$ are magnetoelastic coupling coefficients. By adopting the
short-wave length limit of the magnetostatic dipolar interaction
$\mathcal{H}_{\mathrm{dip}}\approx(\mu_{0}M_{0}^{2}/2)m_{y}^{2}\sin^{2}%
\theta,$ we disregard the Damon-Eshbach surface modes \cite{Damon1961} and
simplify the dispersion of the volume modes, which is allowed for small laser
spot sizes with response being dominated by high-momentum wave vectors
\cite{AR2012}.

By introducing the forces and torques $\mathbf{F}$ acting on the displacement
vector $\boldsymbol{\Phi}=(m_{y},m_{z},R_{l},R_{t},R_{z})^{T}$, one can write
out the linearized equations of motion as shown in the Supplemental Material~\cite{Supplement}.
Here, the lattice displacement is rewritten in the form of longitudinal
($R_{l}$), in-plane transverse ($R_{t}$) and out-of-plane transverse ($R_{z}$)
modes. Strictly speaking, the damping of phonons and magnon are not
necessarily independent, since magnetization is affected by phonon attenuation
via the MEC \cite{Vittoria10}. We treat Gilbert damping constant $\alpha$ and
phonon relaxation time $\tau_{p}$ as independent parameters since Gilbert
damping can also be caused by magnetic disorder, surface roughness or
defects~\cite{Siu01}. We define the anisotropic spin wave frequency
$\Omega_{0}=\gamma\mu_{0}\sqrt{H(H+M_{0}\sin^{2}\theta)}$ and the MEC
frequency parameter $\Delta(k)=\sqrt{\gamma b^{2}k^{2}/(4M_{0}\rho\Omega_{0}%
)}$ with $\theta$ being the angle between magnetic field and in-plane wave
vector $\mathbf{k}$.

The spatiotemporal dynamics of $\boldsymbol{\Phi}(\mathbf{r},t)$ reads
\begin{equation}
\Phi_{i}(\mathbf{r},t)=\int d\mathbf{r}^{\prime}dt^{\prime}G_{ij}%
(\mathbf{r}-\mathbf{r}^{\prime},t-t^{\prime})F_{j}(\mathbf{r}^{\prime
},t^{\prime}), \label{DyPhi}%
\end{equation}
where $G_{ij}$ are the components of the Green function matrix (propagator)
associated with the magnetoelastic equations of motion specified in the
Supplement~\cite{Supplement}. A femtosecond laser pulse generates forces via the inverse Faraday
effect~\cite{Kirilyuk10,Reid10,Satoh12} and
heating~\cite{Wright92,Rossignol05,Park05,Jager15} that are instantaneous on
the scale of the lattice and magnetization dynamics. The relative importance
of these two mechanisms depends on the material and light and is still a
matter of controversy. Here we find that spot excitation of thin magnetic
films is an appropriate method to separate the two, since they lead to
conspicuous differences in the time and position dependent response.

We consider circularly polarized light along $z$ that by the IFE generates an
effective magnetic field along the same direction. For a femtosecond Gaussian
pulse with spot size $W$, the generated magnetic field has a spatial
distribution $\mathbf{H}_{\mathrm{IFE}}(\mathbf{r},t)=\hat{z}H_{\mathrm{IFE}%
}f(t)\exp(-r^{2}/W^{2})$ where temporal shape $f(t)\approx\tau_{l}\delta(t)$
with pulse duration $\tau_{l}$; the amplitude $H_{\mathrm{IFE}}=\beta
I_{\mathrm{in}}\sigma$ is proportional to laser intensity ($I_{\mathrm{in}}$)
and IFE coefficient ($\beta$), respectively. 
The latter is related to the Verdet constant ($V$) as $\beta=V\lambda_0/(2\pi c_0)$, where $\lambda_0$ and $c_0$ are wavelength and velocity of the light \cite{Pershan66}.
$\sigma=1(-1)$ for left(right)-handed
polarization. The torque $F_{m_{y}}(\mathbf{r},t)=\gamma\tau_{l}\mu
_{0}H_{\mathrm{IFE}}\delta(t)\exp(-r^{2}/W^{2})$. On the other hand, the light
pulse generates a sudden increase of the local lattice temperature $\delta
T(\mathbf{r},t)=[\Gamma I_{\mathrm{in}}\tau_{l}/(\rho C_{v})]\Theta
(t)\exp(-r^{2}/W^{2})$, where $\Gamma$ is the light-absorption coefficient. By
choosing the Heaviside step function $\Theta(t)$ we assume that the lattice
locally equilibrates much faster than the response time of the coherent
magnetization (a few picoseconds \cite{Zijlstra13}), while the subseqent
cooling of the lattice by diffusion is slow. The resulting in-plane
thermoelastic stress $F_{R_{l}}=\eta(3\lambda+2\mu)\partial_{r}T$ generates
longitudinal (pressure) waves~\cite{Davies93,Rossignol05}, where $\eta$ is the
thermoelastic expansion coefficient. The local thermal expansion also
generates a \textquotedblleft bulge\textquotedblright\ shear
stress~\cite{Dewhurst82} at a free surface, i.e., an out-of-plane displacement
$R_{z}$. $F_{R_{z}}=\zeta\eta\nabla^{2}T$, where $\zeta$ is a parameter
proportional to the film thickness and controlled by the substrate and an
eventual cap layer, leads to displacement proportional to local temperature
gradient (see numerical results below). In the Supplemental Material~\cite{Supplement}, we
specify the material parameters for yttrium iron garnet (YIG) adopted in our calculations.

\begin{figure}[ptb]
\includegraphics[width=6.cm]{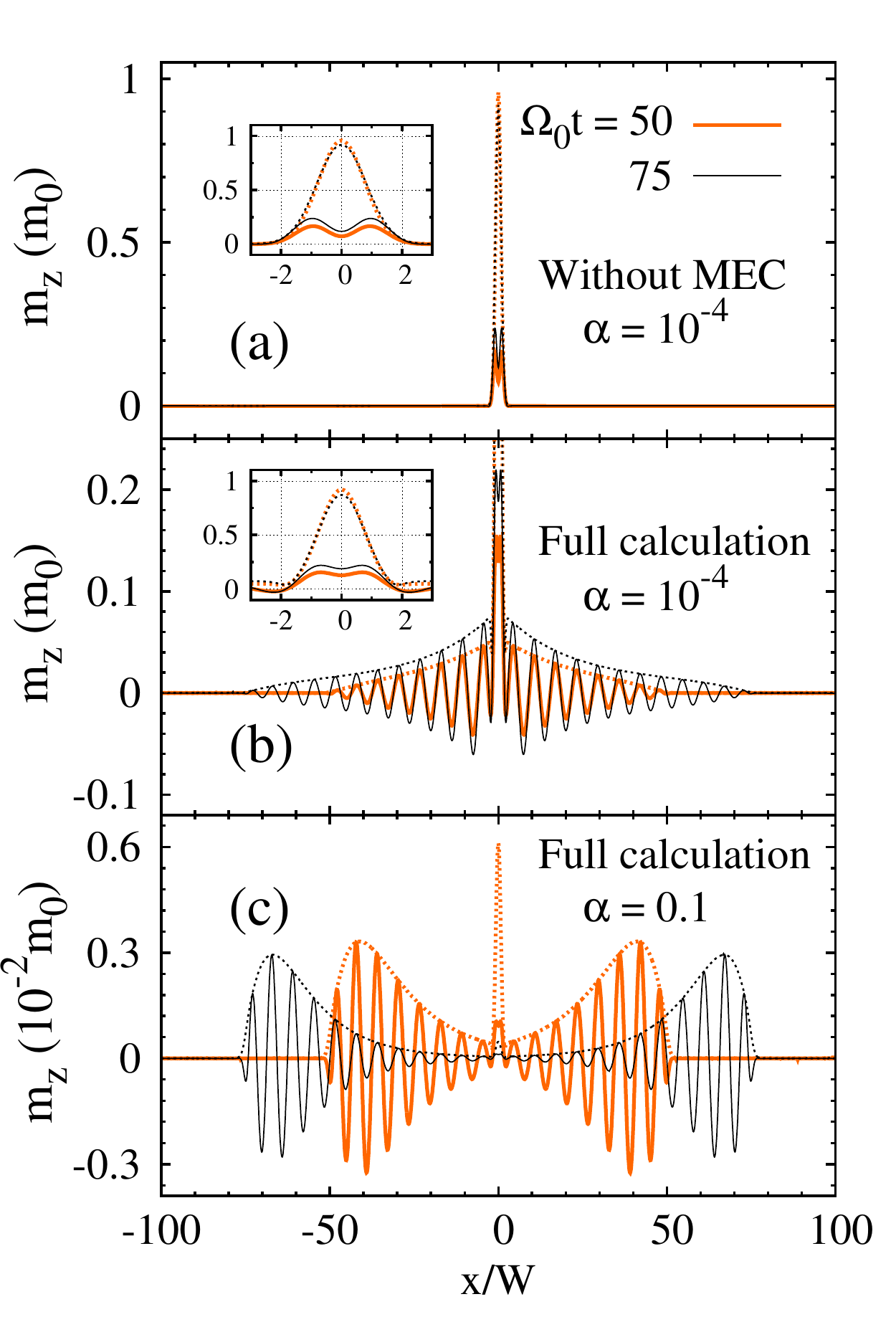}\caption{(Color online)
One-dimensional model for the dynamics of the out-of-plane magnetization
component $m_{z}$ induced by the inverse Faraday effect: (a) At times
$\Omega_{0}t=50$ and $75$ for a Gaussian laser intensity spot in the absence
of MEC with Gilbert damping $\alpha=10^{-4}$. The dashed envelopes are the
modulus ($|\mathbf{m}-\hat{x}|$). \textit{\ } (b) with $\alpha=10^{-4}$ and
(c) with $\alpha=0.1$ are computed for MEC parameter $\tilde{\Delta}(\tilde
{k}_{c})=0.02$. Common are the exchange parameter $\tilde{D}=0.02$, sound
velocity $\tilde{c}_{t}=1$, and sound attenuation rate $\tilde{\tau}_{p}%
^{-1}=10^{-3}$. Note the change of scale between (b) and (c). The insets
provide an expanded view of the laser spot. }%
\label{1Dfigure}%
\end{figure}

\textit{One-dimensional dynamics} --- We start with a spin wave propagating
along the external magnetic field, i.e., $\theta=0$, which by symmetry couples
only with the transverse phonons. The IFE generates the torque $F_{m_{y}%
}(x,t)=m_{0}\delta(t)\exp(-x^{2}/W^{2})$ with $m_{0}=\gamma\tau_{l}\mu
_{0}H_{\mathrm{IFE}}$. This can be realized by a line-shaped excitation
spot~\cite{Satoh12}.

The calculated magnetization profiles at $\Omega_{0}{t}=50$ and $75$ without
and with MEC are plotted in Fig.~\ref{1Dfigure}(a) and (b) separately for
$\alpha=10^{-4}$. Without MEC the magnetization is localized at the exposure
spot and broadens only very weakly with time while the MEC strongly enhances
the broadening of primary magnetization packet, with a wavefront propagating
with the sound velocity ${c}_{t}$. This phenomenon illustrates that the
lattice plays an essential role for spin transport in magnetic films.

Fig.~\ref{1Dfigure}(c) illustrates the sound-assisted propagation for Gilbert
damping $\alpha=0.1$. Instead of the expanding wave front in
Fig.~\ref{1Dfigure}(b), we now find two packets escaping the excitation region
into opposite directions. The packets have a much longer lifetime than the
coherently generated IFE magnetization, hence dominate at long time scales.
This behavior is recovered by the asymptotic expression obtained when
$\alpha\gg\tilde{\Delta}({\tilde k}_{c})$ at the magnon-phonon dispersion crossing wave
vector ($\tilde{k}_{c}=1/\tilde{c}_{t}$),
\begin{align}
m_{z}(\tilde{x},\tilde{t})  &  \simeq m_{0}e^{-\alpha\tilde{t}-\tilde{x}^{2}%
}\sin\left(  \tilde{t}\right)  +(2\alpha)^{-1}m_{0}e^{-\tilde{t}/\tilde{\tau
}_{p}}\tilde{\Delta}^{2}({\tilde k}_{c})\nonumber\\
&  \hspace{-1.2cm}\times%
\begin{cases}
{2\alpha}\tilde{c}_{t}^{2}\left[  \Lambda_{1}(\tilde{c}_{t}\tilde{t}%
-{\tilde{x}})+\Lambda_{1}(\tilde{c}_{t}\tilde{t}+{\tilde{x}})\right]  , &
\tilde{c}_{t}\ll1,\\
\tilde{c}_{t}^{-1}\sqrt{\pi}\left[  \Lambda_{2}(\tilde{c}_{t}\tilde{t}%
-\tilde{x})+\Lambda_{2}(\tilde{c}_{t}\tilde{t}+\tilde{x})\right]  , &
\tilde{c}_{t}\geq1,
\end{cases}
\label{weakasy}%
\end{align}
where $\Lambda_{1}({\xi})=(1/\sqrt{\pi})\int_{0}^{\infty}d\tilde{k}\tilde
{k}^{2}\sin(\tilde{k}{\xi})\exp(-\tilde{k}^{2}/4)$ and $\Lambda_{2}({\xi
})=\exp\left[  -({\xi}^{2}\alpha^{2}+1)/(2\tilde{c}_{t})^{2}\right]  \sin
({\xi}/\tilde{c}_{t})$ and $W$ and $\Omega_{0}$ have been by rendered
dimensionless as explaind in the Supplemental Material~\cite{Supplement}. The (purely magnetic) first term on the
right-hand side represent the exponential decay of the initially excited wave
packet, while the second term is a propagating MP mode. The latter decays with
the phonon damping rate, hence may have a very long mean free path for
materials with high acoustic quality like YIG (assuming that doping affects
the magnetization without increasing sound attenuation).

When the laser spot size is large relative to the MP wave length, i.e.,
$\tilde{c}_{t}\ll1$, according to Eq.~(\ref{weakasy}) the ratio between MP
amplitude and IFE strength scales as $\tilde{c}_{t}^{2}\tilde{\Delta}%
^{2}({\tilde k}_{c})$, i.e., increases with sound velocity and decreases with spot
size. In the other limit, $\tilde{c}_{t}\gg1$, the amplitude of the long-lived
signal is inversely proportional to $\tilde{c}_{t}$, therefore decreases with
increasing $\tilde{c}_{t}$. We therefore estimate this ratio to be maximal
$e^{-1/4}{\sqrt{\pi}\tilde{\Delta}^{2}({\tilde k}_{c})}/({2\alpha})$ when the laser
spot size matches the MP wave length. The peak amplitude of MPs in
Fig.~\ref{1Dfigure}(c) is around $3\times10^{-3}$, in good agreement with
$e^{-1/4}{\sqrt{\pi}\tilde{\Delta}^{2}({\tilde k}_{c})}/({2\alpha})\approx
2.7\times10^{-3}$. \begin{figure}[ptb]
\includegraphics[width=6cm]{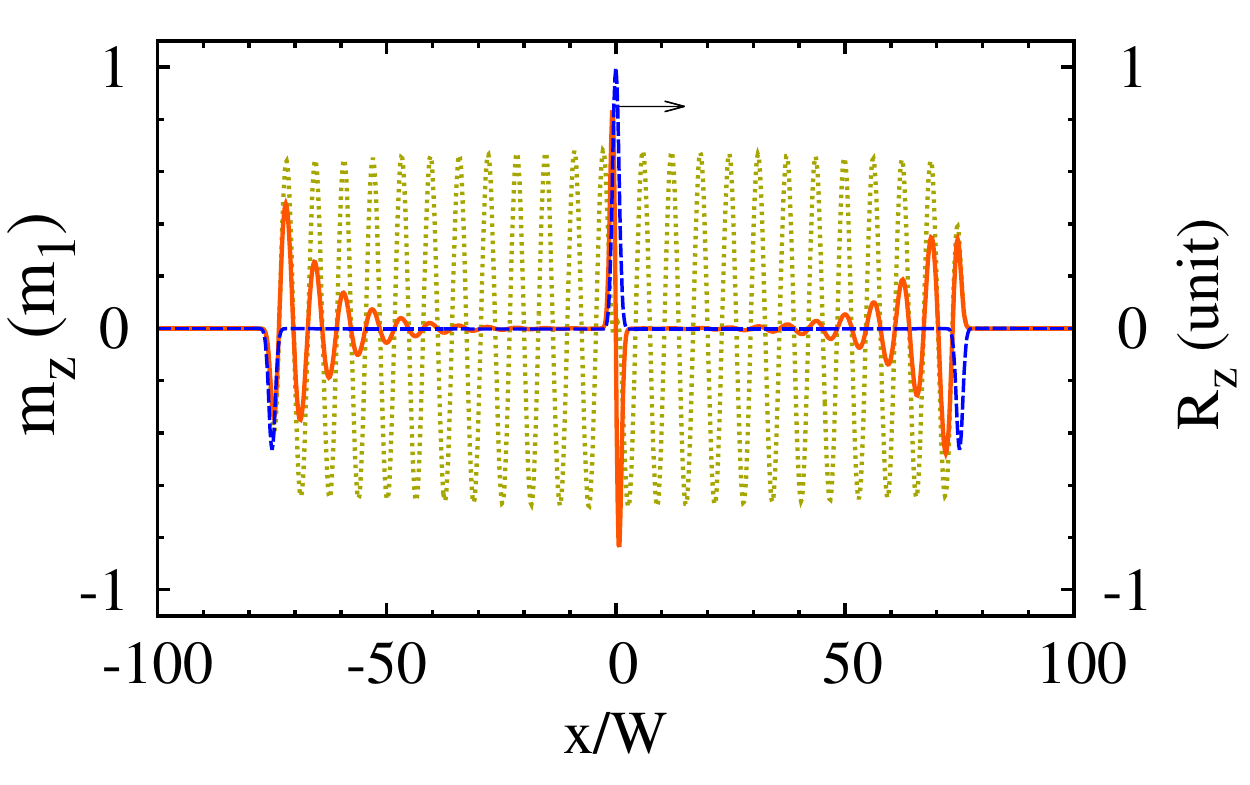}\caption{(Color online) Out-of-plane
magnetization dynamics $m_{z}$ with Gilbert damping $\alpha=10^{-4}$ (dotted
curve) and $0.1$ (solid curve)\ induced by the spot heating by a laser pulse.
The blue dashed curve shows the displacement profile ($R_{z}$). Other
parameters are those in Fig.\thinspace\ref{1Dfigure}. }%
\label{1Dfigureheat}%
\end{figure}

Thermal actuation is caused by the shear force generated by the laser heating
profile $F_{R_{z}}(x,t)=(\zeta\eta c_{t}^{2}\Gamma I_{\mathrm{in}}\tau
_{l}/C_{v})\Theta(t)\partial_{x}^{2}\exp(-x^{2}/W^{2})$, since the pressure
wave is decoupled from the spin wave at $\theta=0$. The asymptotic expression
for $\alpha\gg\tilde{\Delta}({\tilde k}_{c})$ becomes
\begin{align}
m_{z}(x,t)  &  =m_{1}\nonumber\\
&  \times%
\begin{cases}%
\begin{array}
{cc}
\tilde{c}_{t}\{\Lambda_{3}(\tilde{x})-(1/2)e^{-\tilde{t}/\tilde{\tau}_{p}} &
\\
\times\lbrack\Lambda_{3}(\tilde{c}_{t}\tilde{t}+\tilde{x})-\Lambda_{3}%
(\tilde{c}_{t}\tilde{t}-\tilde{x})]\} &
\end{array}
, & \tilde{c}_{t}\ll1,\\%
\begin{array}
{cc}
\tilde{c}_{t}(1-e^{-\alpha t}\cos t)\Lambda_{3}(\tilde{x})+\tilde{c}_{t}%
^{-1}(\sqrt{\pi}/4) & \\
\times e^{-\tilde{t}/\tilde{\tau}_{p}}\left[  \Lambda_{4}(\tilde{c}_{t}%
\tilde{t}+{\tilde{x}})-\Lambda_{4}(\tilde{c}_{t}\tilde{t}-{\tilde{x}})\right]
&
\end{array}
, & \tilde{c}_{t}\geq1,
\end{cases}
\label{Fz_my}%
\end{align}
where $\Lambda_{3}({\xi})=\xi\exp(-\xi^{2})$ and $\Lambda_{4}({\xi}%
)=\exp[-({\xi}^{2}\alpha^{2}+1)/(2\tilde{c}_{t})^{2}]\cos({\xi}/\tilde{c}%
_{t})$. For YIG, the parameter $m_{1}=\gamma b\zeta\eta\Gamma I_{\mathrm{in}%
}\tau_{l}/(2M_{0}\rho c_{t}C_{v})\sim10^{3}\zeta\delta T\,\mathrm{m}%
^{-1}\mathrm{K}^{-1}$. Compared to Eq.~(\ref{weakasy}), the heat-induced
magnetization has (\textit{i}) odd parity in real space, i.e., $m_{z}%
(-x,t)=-m_{z}(x,t)$, (\textit{ii}) a long-lived localized signal near the
excitation spot, and (\textit{iii}) maximum amplitude of propagation at
$\tilde{c}_{t}\simeq1$, cf. Fig.~\ref{1Dfigureheat}. We also plot the
amplitude of the thermally generated phonon wave front that is trailed by the magnetization.

\begin{figure}[ptb]
\includegraphics[width=8cm]{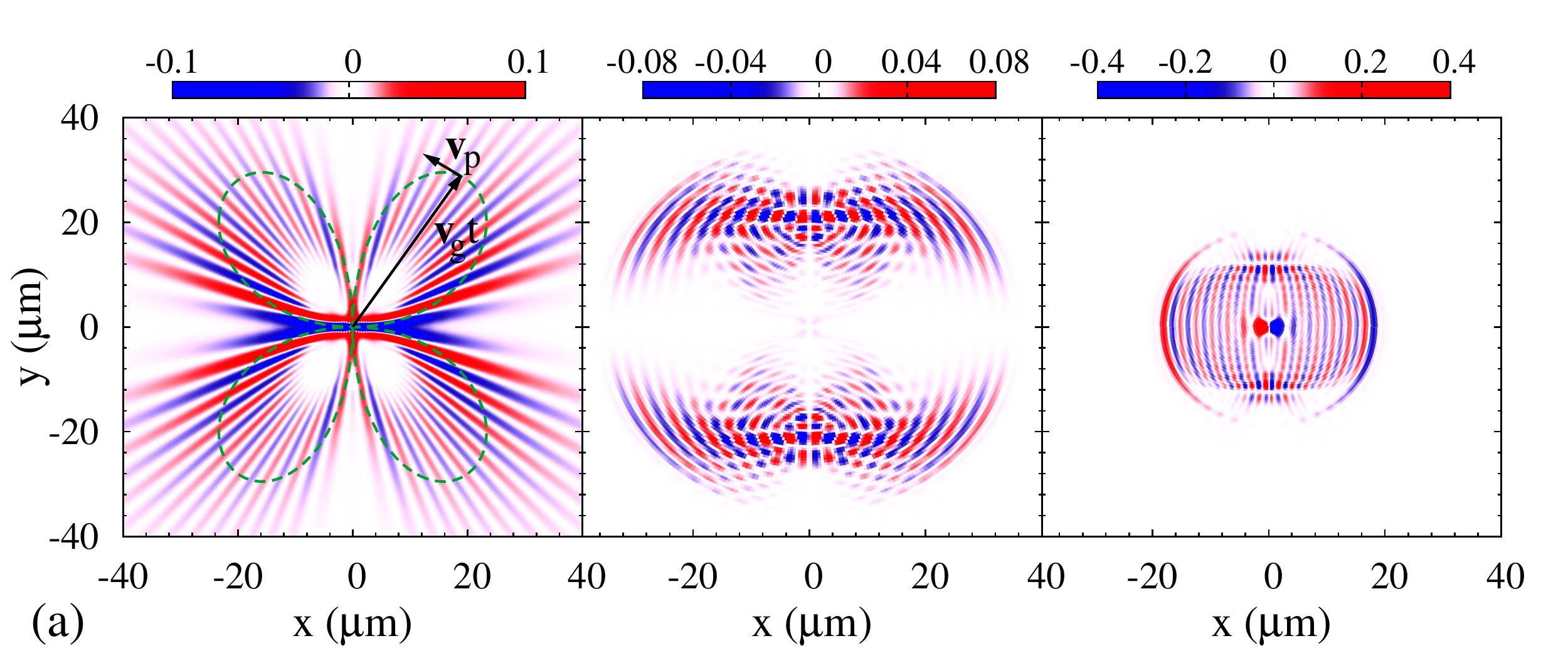}
\includegraphics[width=8cm]{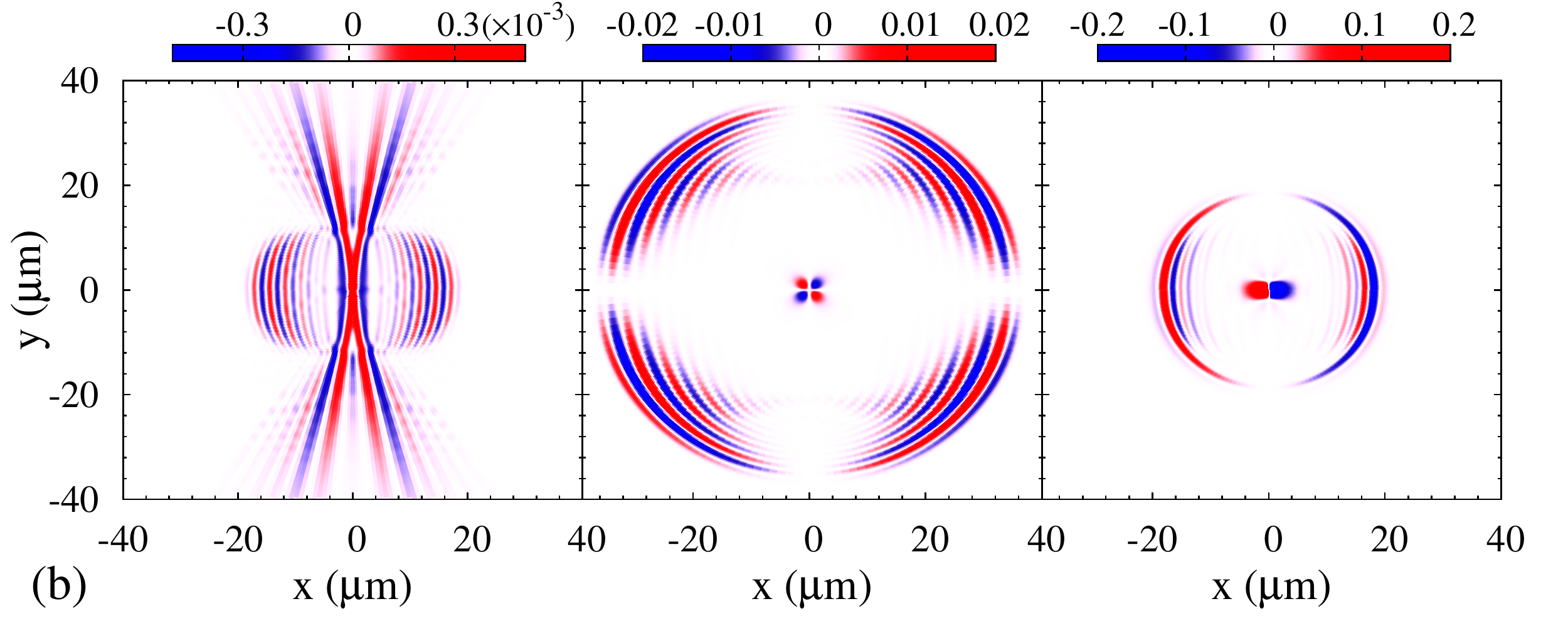}\caption{(Color online) Two
dimensional profile of out-of-plane magnetization, $m_{z}$, at $t={5\,}%
$\thinspace{ns} due to (left) IFE field $F_{m_{y}}$, (middle) pressure stress
$F_{R_{l}}$ and (right) shear stress $F_{R_{z}}$. We normalize the result by
$m_{0}$ for (left), $m_{2}=\gamma b\eta\Gamma I_{\mathrm{in}}\tau_{l}%
(3\lambda+2\mu)/(2M_{0}\rho^{2}C_{v}c_{t}^{2}\Omega_{0})$ for (middle) and
$m_{1}$ for (right). The Gilbert damping coefficient $\alpha=10^{-4}$ and 0.1
for (a) and (b), respectively. The dashed (olive) curve in (a, left)
illustrates the angular-dependent spin wave group velocity. }%
\label{2Dfigure}%
\end{figure}

\textit{Two Dimensions} --- In the following, magnetization is oriented along
$\hat{x}$ by an external in-plane magnetic field $\mu_{0}H={50}\,\mathrm{mT}$
corresponding to $\gamma\mu_{0}H/(2\pi)\simeq{1.4}\,\mathrm{GHz.}$ The spot
size $W={1}\,\mathrm{\mu m}$ and the dimensionless velocities are $\tilde
{c}_{t}\approx0.43$ and $\tilde{c}_{l}\approx0.82$. Fig.~\ref{2Dfigure}
summarizes our main results for the IFE and heat induced dynamics in terms of
the out-of-plane magnetization component $m_{z}$. We plot a snapshot at
$t={5}$\thinspace{ns} in the $x$-$y$ (film) plane from the calculation with
low ($\alpha=10^{-4}$) and enhanced ($\alpha=0.1$) magnetic damping in (a) and
(b), respectively. Fig.~\ref{2Dfigure}(a, left) displays IFE actuated outgoing
rays that broaden with distance from the excitation spot. This feature is
insensitive to MEC strength and can be understood by the angular dependent
group velocities of magnetostatic spin wave dispersion around the average
modulus of the wave vectors $k_{0}$. As discussed in the Supplemental
Material~\cite{Supplement}, the group velocity $\mathbf{v}_{g}\simeq\hat{\theta}[\gamma\mu
_{0}M_{0}/(2k_{0})]\sin2\theta\sqrt{H/(H+M_{0}\sin^{2}\theta)}$ generates an
expansion of the initial wave packet as shown by the dashed (olive) curve,
while the the star-like interference fringes are governed by the phase
velocity. At larger magnetic damping, cf. Fig.~\ref{2Dfigure}(b, left), the
star-like features in the $\hat{x}$ direction are suppressed in favor of
MP\ propagation with transverse sound velocity $c_{t}$. Dotted feature around
$c_{l}t$ are caused by interference of the longitudinal MP and the damped
residue of the initial magnetization wave packet with $\theta\simeq0$, which
has relative longer lifetime. Note the mirror symmetry with respect to the $y$
axis, $m_{z}\left(x,y\right)  =m_{z}\left(  -x,y\right)  $%
.

The quadrupolar features in Figs.~\ref{2Dfigure}(middle) with nodes along the
$x$ and $y$ axes and $\sin2\theta$ symmetry are induced by the pressure
$F_{R_{l}}$ caused by a heat pulse. The radii of the circular wave fronts
correspond to the longitudinal sound velocity. Figs.~\ref{2Dfigure}(right)
illustrate that a shear stress induces MPs that spread with transverse sound
velocity $c_{t},$which are thereby clearly distinguishable from the pressure
induced signals: $F_{R_{z}}$ generates dipolar symmetric features with nodes
at the $y$ axis, which follows from the $\cos\theta$ symmetry of the MEC
coupling. Clearly, both heat-induced signals are antisymmetric with respect to
reflection at the $y$ axis, $m_{z}(x,y)=-m_{z}(-x,y)$, which allows
discrimination from the IFE response. Moreover, we identify a non-propagating
signal in the vicinity of the excitation spot [see the center of
Figs.~\ref{2Dfigure}(b, middle) and (b, right)], a \textquotedblleft smoking
gun\textquotedblright\ for thermally excited dynamics.

It is not easy to predict the absolute and relative magnitude of the two
mechanisms for a given light intensity from first principles due to
uncertainties in the strongly non-equilibrium processes after an intense fs
light pulse. Microscopic theories address the ultra-fast physics of angular
momentum and energy transfer from the light to the magnetic order
\cite{Lefkidis09,Atxitia09} and the lattice \cite{Zijlstra13} and should
ideally be employed to fix the initial conditions for our calculations. But
also the long-time response depends on several temperature and frequency
dependent materials parameters that govern the IFE, light absorption etc.
Satoh \textit{et al.}~\cite{Satoh12} find a Faraday rotation of the probe
pulse of the order of milli-rad for 110~$\mathrm{\mu}$m thick bismuth-doped
iron garnet, which corresponds to a light-induced torque of
$m_{0}\sim 0.004$ for a Verdet constant of 
$10^{4}$ \textrm{rad\thinspace m}$^{-1}$\textrm{T}$^{-1}$%
\textrm{\ }\cite{Supplement}. With thermal expansion coefficient $\eta\sim10^{-5}%
$/K~\cite{Boudier04}, the thermal torque $m_{2}\sim4\times10^{-3}\delta T$/K,
which can be larger than $m_{0}$ for pulsed laser-induced
heating~\cite{Choi15}. Similar values may be anticipated for $m_{1}$ when
effective thicknesses $\zeta\sim\mathrm{\mu}$m. We should also not forget 
that the
fast light-induced demagnetization \cite{Beaurepaire96} should affect the
response directly under the excitation spot, but its diffusion should be
slower than the ballistic response computed here.

\textit{Conclusion and Discussion} --- We modeled the spatiotemporal
laser-induced magnetization dynamics in magnetic thin films, concluding that
magnetoelastic coupling is essential for spin angular momentum transport
because the phonon group velocity is much larger than that of the magnons. An
experimental study of the symmetry of time-resolved magnetization wave fronts
radiating from the excitation spot allows discriminating different laser
excitation mechanisms, thereby helping to answer the long-standing question on
the physical origin of ultra-fast magnetization dynamics, i.e. whether it is
caused by coherent light-induced magnetic fields or sudden heating of the
lattice. Moreover, we clarified the optimal size of the excitation laser spot
to be the MP wavelength; for YIG at an applied field of $50$~mT it is
$\sim1\,\mathrm{\mu m}$.

The essential role of the MEC coupling might have larger ramifications. For
example, a number of recent experiments on the spin Seebeck effect on YIG came
to the conclusion that the thermal spin pumping is not caused by\ terhahertz
magnons at energies around $k_{B}T$, but by spin waves in a low energy band
close to the gap~\cite{Kikkawa15,Jin15,Guo15}. Spin information was found to
propagate in YIG diffusely over large distances~\cite{Cornelissen15,Giles15}.
From the present results we venture that strongly coupled magnon-polarons
could be the carriers of thermal spin currents.

This work is supported by the DFG Priority Program 1538 SpinCat, the FOM
foundation, the E-IMR and ICC-IMR, and the JSPS (Grant Nos. 25247056, 25220910, 26103006). We
acknowledge discussions with Benedetta Flebus and Rembert Duine.

Note added in proof: Ogawa \textit{et al.} report generation of laser-generated magnon polarons that drive magnetic bubble domains \cite{Ogawa15}.
\bibliographystyle{prsty}
\bibliography{Refs.bib}

\end{document}


\title{Supplementary material to \textquotedblleft Laser-induced spatiotemporal
dynamics of magnetic films"}
\author{Ka Shen}
\affiliation{Kavli Institute of NanoScience, Delft University of Technology, Lorentzweg 1,
2628 CJ Delft, The Netherlands}
\author{Gerrit E. W. Bauer}
\affiliation{Institute for Materials Research and WPI-AIMR, Tohoku University, Sendai
980-8577, Japan}
\affiliation{Kavli Institute of NanoScience, Delft University of Technology, Lorentzweg 1,
2628 CJ Delft, The Netherlands}
\maketitle

\makeatletter
\makeatother

\setcounter{table}{0} \renewcommand{\thetable}{S\arabic{table}}
\setcounter{figure}{0} \renewcommand{\thefigure}{S\arabic{figure}}
\setcounter{equation}{0} \renewcommand{\theequation}{S\arabic{equation}} This
supplement includes the linearized equations of motion in Fourier space with
choice of parameters, as well as analytical expression for the IFE-induced
dynamical response to line and point sources and estimation on the magnitude
of IFE.

\section{1. Linearized equations of motion in Fourier space}

After introducing the forces and torques $\mathbf{F}$ acting on the
displacement vector $\boldsymbol{\Phi}=(m_{y},m_{z},R_{l},R_{t},R_{z})^{T}$,
the linearized equations of motion in Fourier space read~\cite{Schlomann60}
\begin{align}
(\omega^{2}-2i\omega\tau_{p}^{-1}-c_{t}^{2}k^{2})\rho R_{t}  &  =ibk\cos
2\theta m_{y}-{F}_{R_{t}},\label{EqRt}\\
(\omega^{2}-2i\omega\tau_{p}^{-1}-c_{t}^{2}k^{2})\rho R_{z}  &  =ibk\cos\theta
m_{z}-F_{R_{z}},\\
(\omega^{2}-2i\omega\tau_{p}^{-1}-c_{l}^{2}k^{2})\rho R_{l}  &  =ibk\sin
2\theta m_{y}-F_{R_{l}},\label{EqRl}\\
i\omega m_{y}+(\gamma H_{k}+i\alpha\omega)m_{z}  &  =i\Delta_{1}\cos\theta
R_{z}+F_{m_{y}},\label{Eqmy}\\
i\omega m_{z}-(\gamma H_{k}^{\prime}+i\alpha\omega)m_{y}  &  =-i\Delta
_{1}(\cos2\theta R_{t}+\sin2\theta R_{l}) +F_{m_{z}}, \label{Eqmz}%
\end{align}
where $\Delta_{1}=\gamma bk/M_{0}$, $H_{k}^{\prime}=H_{k}+\mu_{0}M_{0}\sin
^{2}\theta$ with $H_{k}=\mu_{0}H+\gamma^{-1}(2A/M_{0})k^{2},$ and $\theta$ is
the angle between magnetic field and in-plane wave vector $\mathbf{k}$.
$\alpha$ and $\tau_{p}$ are the Gilbert damping constant and phonon relaxation
time~\cite{Kittel58}.

We adopt the material parameters of yttrium iron garnet (YIG). Specifically,
the dipolar field and gyromagnetic ratio are $\mu_{0}M_{0}={175\,}\mathrm{mT}$
and $\gamma/2\pi=2.8\times10^{10}\,\mathrm{Hz/T}$,
respectively~\cite{Manuilov09}. The MEC constant $b={6.96\times10}%
^{5}\,\mathrm{J/m}^{3}$~\cite{Gurevich96}, the exchange parameter
$2A/M_{0}=9.2\times10^{-6}\,\mathrm{m}^{2}\mathrm{/s}$~\cite{Serga10}, and the
mass density $\rho={5170}\,\mathrm{kg/m}^{3}$~\cite{Gilleo58}. The transverse
and longitudinal sound velocities are $c_{t}=3.8\times10^{3}\,\mathrm{m/s}$
and $c_{l}=7.2\times10^{3}\,\mathrm{m/s}$~\cite{Gurevich96}. The phonon
relaxation rate in YIG is of the order of ${1}\,\mathrm{\mu s}^{-1}%
$~\cite{Lewis68}. In high quality films, the Gilbert damping coefficient is
typically around $10^{-4}$~\cite{Manuilov09}. For comparison, we also show
results for $\alpha=0.1$, which is close to the upper limit for Bi-doped
YIG~\cite{Siu01}.

\section{2. Analytical expression for IFE-induced one-dimensional dynamics}

The time evolution of the out-of-plane magnetization Eq.~(2) of the main text
can be simplified to
\[
m_{z}(\tilde{x},\tilde{t})=\frac{m_{0}}{\sqrt{\pi}}\sum_{\pm}\int_{0}^{\infty
}d\tilde{k}e^{-\tilde{k}^{2}/4}\operatorname{Im}\left(  \frac{\tilde{\omega
}_{\mp}-\tilde{\Omega}_{0}}{\tilde{\omega}_{\mp}-\tilde{\omega}_{\pm}%
}e^{i\tilde{\omega}_{\pm}\tilde{t}}\right)  \cos\tilde{k}\tilde{x},
\]
where $\omega_{\pm}=\left[  \tilde{\Omega}_{0}+\tilde{\Omega}_{p}\pm
\sqrt{(\tilde{\Omega}_{0}-\tilde{\Omega}_{p})^{2}+8\tilde{\Delta}^{2}}\right]
/2$. Here $\tilde{\Omega}_{0}=(1+i\alpha)(1+\tilde{D}\tilde{k}^{2})$,
$\tilde{\Omega}_{p}=\tilde{c}_{t}\tilde{k}+i\tilde{\tau}_{p}^{-1}$, $\tilde
{k}=kW$, $\tilde{x}=x/W$, $\tilde{\tau}_{p}=\Omega_{0}\tau_{p}$,
$\tilde{\Delta}=\Delta/\Omega_{0}$ $\tilde{c}_{t}=c_{t}/(W\Omega_{0})$,
$\tilde{t}=\Omega_{0}t$, and $\tilde{D}=2A/(M_{0}\Omega_{0}W^{2})$ are all dimensionless.

\section{3. Analysis of star-like feature due to IFE from point sources}

\begin{figure}[ptb]
\includegraphics[width=6.7cm]{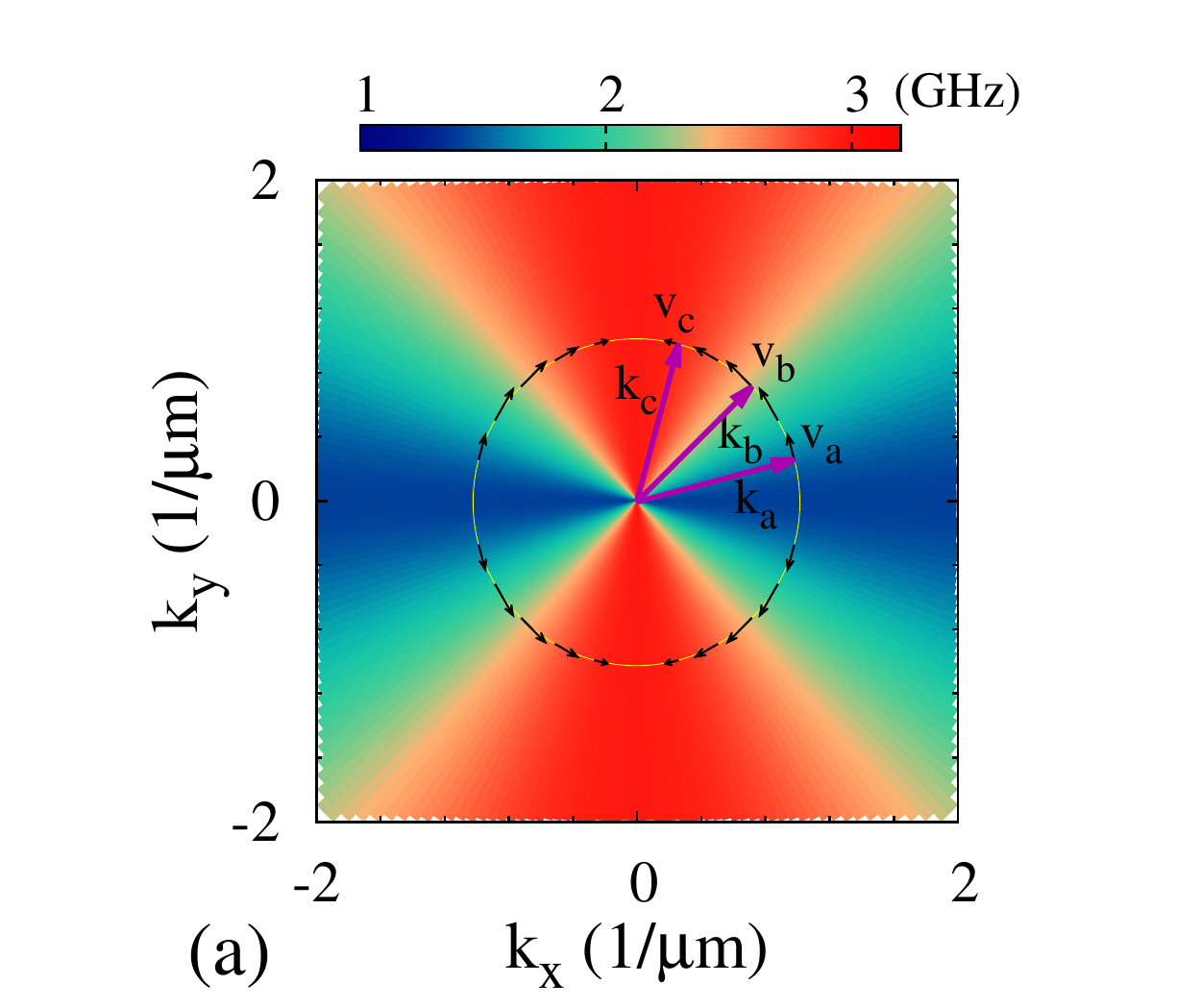}
\includegraphics[width=6.cm]{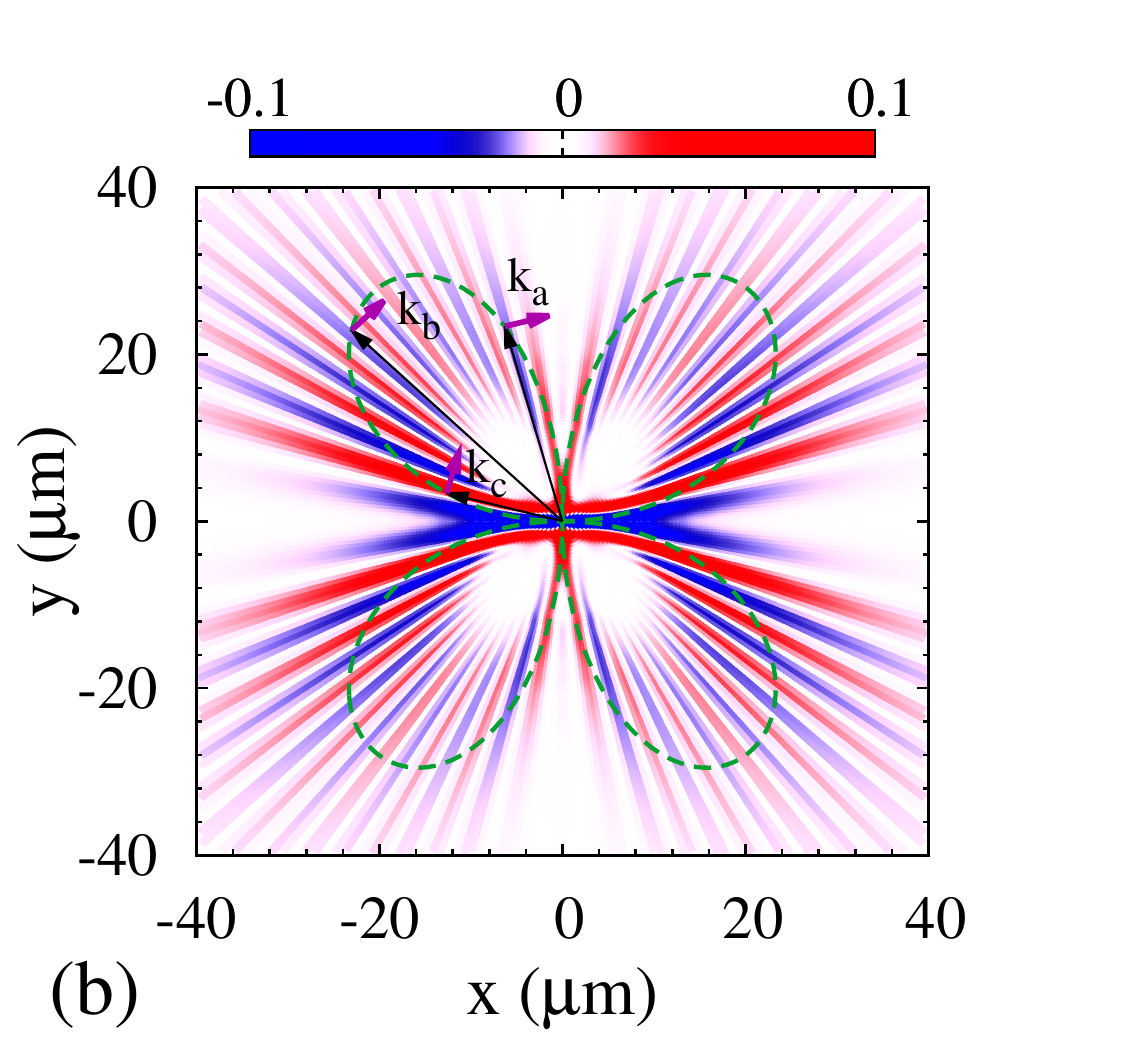} \caption{(Color online) (a)
Approximate (bulk) spin wave spectrum $\omega_{\mathbf{k}}=\gamma\mu_{0}%
\sqrt{H(H+M_{0}\sin^{2}\theta)}$ in $\mathbf{k}$-space. The thin (black)
arrows show angular dependence of the group velocities $\mathbf{v}_{g}%
=\nabla_{\mathbf{k}}\omega_{\mathbf{k}}$ at the average modulus $k_{0}%
=\langle|\mathbf{k}|\rangle$. Those for three typical wave vectors
$\mathbf{k}_{a,b,c}$ represented by thick (purple) arrows are labeled by
$\mathbf{v}_{a,b,c}$. (b) Replotting of Fig.4(a,left), IFE-induced
magnetization profile in real space at $5$~ns, with three thin (black) [thick
(purple)] arrows representing $t\mathbf{v}_{a,b,c}$ [$\mathbf{k}_{a,b,c}$].}%
\label{S1}%
\end{figure}The IFE-amplitude induced by an axially symmetric light pulse as
plotted in Fig.~4(a,left) displays suprisingly complex interference patterns.
These features can be explained in terms of the angular-dependent group
velocity at constant $k_{0}=\langle|\mathbf{k}|\rangle$ [yellow circle in the
(volume) spin wave dispersion shown in Fig.~\ref{S1}(a)]. The thin (black)
arrows indicate the group velocities $\mathbf{v}_{g}\equiv\nabla_{\mathbf{k}%
}\omega_{\mathbf{k}}|_{k_{0},\theta}\simeq\hat{\theta}[\gamma\mu_{0}%
M_{0}/(2k_{0})]\sin(2\theta)\sqrt{H/(H+M_{0}\sin^{2}\theta)}$, where we used
the simplified dispersion $\omega_{\mathbf{k}}\simeq\gamma\mu_{0}%
\sqrt{H(H+M_{0}\sin^{2}\theta)}$ that is accurate for not too small
$k_{0}\gtrsim1/W.$ In this approximation $\mathbf{v}_{g}\cdot\mathbf{k}=0,$
i.e., the group velocity is purely transverse. The thick (purple) arrows
Fig.~\ref{S1}(a) represent three typical wave vectors $\mathbf{k}_{a,b,c}$,
whose corresponding group velocities are labeled by $\mathbf{v}_{a,b,c}$
respectively. In Fig.~\ref{S1}(b), we replot the IFE-induced magnetization
profile in real space with dashed (olive) curve representing $t\left\vert
\mathbf{v}_{g}\right\vert $ and the vectors $\mathbf{k}_{a,b,c}$ and
$t\mathbf{v}_{a,b,c}$ as thick (purple) and thin (black) arrows for
$t=5\,\mathrm{ns}$. We can understand this pattern in terms of the time
dependence of a linear superposition of wave packets arranged around the
origin: Each wave packet centered around a certain $\mathbf{k}_{0}$ is
modulated in the direction ${\hat{k}}_{0}$ with wavelength $1/k_{0}$. The
whole package moves rigidly in the transverse direction with group velocity
$\mathbf{v}_{g}$ that is finite by the dipolar interaction. The initial wave
packet at the origin thereby blows up like $\left\vert t\mathbf{v}%
_{g}\right\vert $. The interference pattern initially normal to $\mathbf{k}%
_{0}$ is in real space shifted and combines with the patterns generated by
other wave packets on the ring to form complex features that in the snapshot
plotted in the figures look like a star. Moreover, the lifetime of spin waves
is determined by $1/(\alpha\omega_{\mathbf{k}})$, leading to $\tau
_{\mathbf{k}_{a}}>\tau_{\mathbf{k}_{b}}>\tau_{\mathbf{k}_{c}}$. This explains
that even for large dampings a radial pattern is still visible near the
$y$-axis while that along the $x$-axis has disappeared [see Fig. 4(b,left)].

Analytically, the out-of-plane magnetization at $\mathbf{r}=(r\cos
\varphi,r\sin\varphi)$ with respect to the center of initial spot can be
expressed as
\begin{equation}
m_{z}(r,\varphi,t)\propto\int_{0}^{2\pi}d\theta e^{-\alpha t\gamma\mu_{0}%
\sqrt{H(H+M_{0}\sin^{2}\theta)}}\int_{0}^{\infty}d\tilde{k}\tilde{k}%
e^{-\tilde{k}^{2}/4}\cos\left(  \tilde{k}\tilde{r}\cos(\theta-\varphi
)-t\gamma\mu_{0}\sqrt{H(H+M_{0}\sin^{2}\theta)}\right)  .
\end{equation}
Far from the excitation spot, i.e., $\tilde{r}\gg1$, the $\tilde{k}$-integral
is dominated by the spin waves satisfying $\cos(\theta-\varphi)\sim0$ or
$\theta\sim\varphi\pm\pi/2$. Taylor-expanding in the phase factor of the
cosine leads to
\begin{align}
m_{z}(r,\varphi,t)  &  \sim e^{-\alpha t\gamma\mu_{0}\sqrt{H(H+M_{0}\cos
^{2}\varphi)}}\int_{0}^{\infty}d\tilde{k}\tilde{k}e^{-\tilde{k}^{2}/4}%
\int_{-\delta\theta}^{\delta\theta}d\theta^{\prime}\cos\left(  \tilde
{k}(\tilde{r}-t|v_{g}|)\theta^{\prime}-t\gamma\mu_{0}\sqrt{H(H+M_{0}\cos
^{2}\varphi)}\right) \nonumber\\
&  \sim2\cos\left(  t\gamma\mu_{0}\sqrt{H(H+M_{0}\cos^{2}\varphi)}\right)
e^{-\alpha t\gamma\mu_{0}\sqrt{H(H+M_{0}\cos^{2}\varphi)}}\int_{0}^{\infty
}d\tilde{k}\frac{\sin\left(  \tilde{k}(\tilde{r}-t|v_{g}|)\delta\theta\right)
}{\tilde{r}-t|v_{g}|}e^{-\tilde{k}^{2}/4} \label{eqx}%
\end{align}
where $\delta\theta$ is a small cut-off angle and $|v_{g}|=(\gamma\mu_{0}%
M_{0}/{\tilde{k}})|\sin2\varphi|\sqrt{H/(H+M_{0}\cos^{2}\varphi)}$.
Eq.~(\ref{eqx}) catches most features observed in Figs. 4(left) as summarized
above: (\textit{i}) $\cos\left(  t\gamma\mu_{0}\sqrt{H(H+M_{0}\cos^{2}%
\varphi)}\right)  $ is an angle-dependent oscillations, i.e., reproduces the
star-like feature. (\textit{ii}) $\exp[-\alpha t\gamma\mu_{0}\sqrt
{H(H+M_{0}\cos^{2}\varphi)}]$ explains the angular dependent attenuation,
revealing that rays with $\varphi\simeq\pi/2$ decay slower than those for
$\varphi\simeq0$. At large damping a radial pattern is visible near the $y$
but not in the $x$-direction [see Fig. 4(b,left)]. (\textit{iii}) The radial
integral is a function of $\tilde{r}-t|v_{g}|$ only, therefore predicts that
the wave packet expands with the group velocity $|v_{g}|$.

\section{4. Magnitude of IFE}

Here we estimate the magnitude of the IFE based on the experiments by Satoh
\textit{et al. }\cite{Satoh12}, who excited spin waves in a 110$\,\mathrm{\mu
}$m thick $\mathrm{Gd}_{4/3}\mathrm{Yb}_{2/3}\mathrm{BiFe}_{5}\mathrm{O}_{12}$
film by laser light with pump fluence of 200~mJ(cm)$^{-2}$ and pulse width of
120~fs at wavelength $1400\,$nm, and measured Faraday rotations of a linearily
polarized probe pulse (at wave length $792\,$nm) of the order of
$\vartheta_{\mathrm{F}}\sim1$~mrad. Satoh \textit{et al.} do not measure the
Verdet constant. In Bi-YIG (with unspecified doping concentration) the Verdet
constants are $V\sim5800$, 1050, and 380~rad\thinspace m$^{-1}$T$^{-1}$ for
wavelengths 543nm, 633nm, and 780nm, respectively \cite{Bai03}. The Verdet
constant in $(\mathrm{BiYbTb})_{3}\mathrm{Fe}_{5}\mathrm{O}_{12}$ at $780\,$nm
was found to be $V\sim8.5\times10^{4}$~rad\thinspace m$^{-1}$T$^{-1}$
\cite{Bai03}. Here we adopt $V=10^{4}$~\textrm{rad\thinspace m}$^{-1}%
$\textrm{T}$^{-1}$ for both pump and probe fields.

The IFE\ and FE\ may be formulated in terms of a phenomenological free
energy~\cite{Pershan66,Kirilyuk10}
\begin{equation}
F=\beta_{\mathrm{pump}}(\mu_{0}M_{z})I_{\mathrm{in}},\mathit{\ }%
\end{equation}
where\ $\beta_{\mathrm{pump}}$\ is\ a\ magneto-optical\ susceptibility at the
pump laser frequency and $I_{\mathrm{in}}$ is the intensity of (right-hand)
circularly polarized light. The effective field $H_{\mathrm{IFE}}=-(1/\mu
_{0})\partial_{M_{z}}F=-\beta_{\mathrm{pump}}I_{\mathrm{in}}.$ In the same
theoretical framework, the Faraday rotation induced in the external field-free
limit by $B=\mu_{0}M_{z}$ reads%
\begin{equation}
\vartheta_{\mathrm{F}}=\frac{2\pi c_{0}}{\lambda_{0}}\beta_{\mathrm{probe}%
}(\mu_{0}M_{z})L=V_{\mathrm{probe}}(\mu_{0}M_{z})L,\label{FarAngle}%
\end{equation}
where $c_{0}$, $\lambda_{0}$, and $L$ are the velocity of light in vacuum, the
wavelength of the probe pulse, and the film thickness, respectively. This
relation allow us to express $\beta$ in terms of $V.$

We can now estimate (assuming $V_{\mathrm{pump}}\approx V_{\mathrm{probe}})$
$|\mu_{0}H_{\mathrm{IFE}}|=\mu_{0}V\lambda_{0}I_{\mathrm{in}}/(2\pi c_{0}%
)\sim0.2$~T for the experimental pump intensity. $H_{\mathrm{IFE}}$ when
applied as a pulse with width $\tau_{l}$ generates a magnetization
$M_{y}=\gamma\tau_{l}\mu_{0}H_{\mathrm{IFE}}M_{0}$ and $m_{0}=M_{y}/M_{0}%
\sim0.004$. Disregarding damping, the precession in the external field
$H\hat{x}$ preserves this amplitude such that $M_{z}\approx M_{y}.$ This leads
to $\vartheta_{\mathrm{F}}\approx V(\gamma\tau_{l}\mu_{0}^{2}H_{\mathrm{IFE}%
}M_{0})L\sim0.5$ mrad, in reasonable agreement with the observations
\cite{Satoh12}.
